\documentclass[final,1p,times]{elsarticle}

\usepackage{epsfig}

\usepackage{amssymb}

\journal{Nuclear Physics B}

\begin{document}

\begin{frontmatter}



\title{Graphene wormholes\\ A condensed matter illustration of Dirac fermions in curved space}


\author{J. Gonz\'alez\corref{cor1}}
\ead{gonzalez@iem.cfmac.csic.es}
\cortext[cor1]{Corresponding author}

\author{J. Herrero}

\address{Instituto de Estructura de la Materia,
        Consejo Superior de Investigaciones Cient\'{\i}ficas, Serrano 123,
        28006 Madrid, Spain}

\begin{abstract}
We study the properties of graphene wormholes in which a short nanotube 
acts as a bridge between two graphene sheets, where the honeycomb carbon 
lattice is curved from the presence of 12 heptagonal defects. By taking
the nanotube bridge with very small length compared to the radius, we develop 
an effective theory of Dirac fermions to account for the low-energy electronic 
properties of the wormholes in the continuum limit, 
where the frustration induced by the heptagonal 
defects is mimicked by a line of fictitious gauge flux attached to each of 
them. We find in particular that, when the effective gauge flux from 
the topological defects becomes maximal, the zero-energy modes of the Dirac 
equation can be arranged into two triplets, that can be thought as the 
counterpart of the two triplets of zero modes that arise in the dual instance 
of the continuum limit of large spherical fullerenes. We further investigate
the graphene wormhole spectra by performing a numerical diagonalization of
tight-binding hamiltonians for very large lattices realizing the wormhole
geometry. 
The correspondence between the number of localized electronic states 
observed in the numerical approach and the effective gauge flux predicted 
in the continuum limit shows that graphene wormholes can be consistently 
described by an effective theory of two Dirac fermion fields in the curved 
geometry of the wormhole, opening the possibility of using real samples of 
the carbon material as a playground to experiment with the interaction 
between the background curvature and the Dirac fields.

\end{abstract}


\end{frontmatter}


\section{Introduction}

In recent years, the feasibility of fabricating in the laboratory individual 
sheets of graphite (so-called graphene) has opened an exciting field of research
in condensed matter physics\cite{novo,geim,kim}. 
This new material illustrates the case of a genuine two-dimensional electron 
system, with remarkable electronic properties and great potential for 
technological applications\cite{rmp}. Many of the interesting features of 
graphene arise actually from the unconventional band structure of the carbon sheet.
The undoped material has conical conduction and valence bands that meet at the 
corners of the hexagonal Brillouin zone, thus mimicking at low energies the 
dispersion of electrons and positrons in the Dirac theory (though with a Fermi 
velocity about 300 times smaller than the speed of light). There have been already 
clear experimental signatures of the Dirac-like dispersion of the electron system
as, for instance, the peculiar sequence of plateaus measured in the quantum Hall 
effect\cite{geim,kim}.
The relativistic invariance of the low-energy theory is also at the origin 
of many of the outstanding properties of graphene, like the insensitivity of 
the  electronic transport to scatterers with size larger than the lattice 
spacing\cite{suzu} or the transparent transmission at normal incidence through 
potential barriers\cite{kats}.

Part of the great attention received by the two-dimensional carbon material has 
come from the possibility of observing phenomena in such an electron system which
otherwise would be confined to the realm of high-energy physics. Under usual
experimental conditions, the interacting theory is in the strong-coupling regime,
what has led to propose effects like the hyper-critical screening of charged 
impurities\cite{fog,shy,per,ter}, 
or quasiparticle features deriving from the renormalization of the 
interaction in the many-body theory\cite{np2,prbr,dsarma}.

In this promising picture of graphene, there are however other aspects that still 
have to be better understood. One of them is the role played by the curvature of
the two-dimensional system. The experimental samples of graphene always show some 
corrugation, that is due in part to the irregularities in the surface of the 
substrate on which they are deposited. However, suspended samples also have ripples, 
which is a clear indication of the natural tendency of graphene to develop a
modulation of the two-dimensional geometry. This effect translates into a small 
spatial dependence of the electron hopping between carbon atoms, that can be 
simulated by the introduction of fictitious gauge fields\cite{rip}. 
These are characterized 
anyhow by having a vanishing flux through the two-dimensional system on the 
average, as long as the graphene sheet does not bend over macroscopic length 
scales.
    
In this paper we will be interested in the effects of the curvature on 
graphene in cases where that is strong enough to modify the topology of the carbon 
lattice. The deviation from a flat two-dimensional geometry over large length scales
implies that the intrinsic curvature cannot vanish everywhere, and this is only 
possible through the appearance of carbon rings that are not six-membered within
the bulk hexagonal lattice of  
the graphene sheet. The most notorious instance where these so-called
topological defects play an important role is provided by the fullerenes, where 
12 pentagonal carbon rings are just enough to close the carbon sheet into a
spherical cage, no matter the size of the molecule\cite{np1}. The carbon lattices
we are going to investigate here are to some extent dual to the 
fullerenes, as the carbon sheet is bent now through the introduction of heptagonal
carbon rings. The two-dimensional geometries have then some content of negative
curvature, describing the matching between different infinite spaces. 
In this respect, the geometry
of a carbon nanotube-graphene junction has been already studied in Ref. \cite{ngj}. 
Here we will focus on the case where the carbon sheet is curved more strongly, to give
rise to a wormhole bridge between two different graphene branches. 

The duality between the fullerene and wormhole geometries becomes more explicit 
after noticing that 12 heptagonal carbon rings are needed to establish the 
connection between the two graphene branches of the wormhole. The presence of the
topological defects has in general important consequences regarding the electronic 
properties of the carbon material. It has been shown that the frustration that they 
induce in the hexagonal carbon lattice can be mimicked by the action of fictitious 
gauge fields on the electron system\cite{np1}. Opposite to the case of the gauge fields 
introduced for the ripples, the fields related to the topological defects imply a 
nonvanishing flux threading each odd-membered carbon ring. This means that the 
structure of the low-lying electronic levels is in general dictated by the number of 
topological defects in the carbon lattice, as illustrated in the case of 
the fullerenes\cite{np1,full}. The number of localized states in the carbon 
nanotube-graphene junctions has also shown to be related to the distribution of 
heptagonal carbon rings at the junction\cite{ngj}.
Here we will show that, in the graphene wormholes with maximal gauge flux, the 
localized states can be arranged into two triplets that are the counterpart of the 
zero-energy modes of the fullerene lattices in the continuum limit.

The content of the paper is distributed as follows. In section 2, we work out a 
simplified continuum model of the wormhole geometry. We review next in section 3 
the correspondence between fictitious gauge fields and odd-membered rings
in the carbon lattice. Section 4 is devoted to set up the effective low-energy 
theory of Dirac fermions interacting with the fictitious gauge field in the 
curved geometry of the wormhole, showing the correlation between the effective
gauge flux and the number of localized states in the electron system. The 
numerical investigation of graphene wormhole spectra is described in section 5,
where a comparison with the results of the continuum field theory is also carried
out. Finally, the conclusions of our work are drawn in section 6.

\section{Wormhole geometry}

A graphene layer can be curved into a carbon nanotube from the presence of 
heptagonal carbon rings at the base of the junction. It can be shown that the number
of heptagonal rings needed is always 6, provided that there are no other topological 
defects\cite{ngj}. 
A graphene wormhole consists of a short nanotube bridging two different 
graphene layers. This can be viewed then as two nanotube-graphene junctions connected 
at the end of each nanotube. We will show that the number of topological defects 
implied by this construction is consistent with the value of 
the Euler characteristic computed for the geometry that arises in the continuum limit 
of the graphene wormhole.

\begin{figure}
\begin{center}
\mbox{
\epsfysize 3.5cm \epsfbox{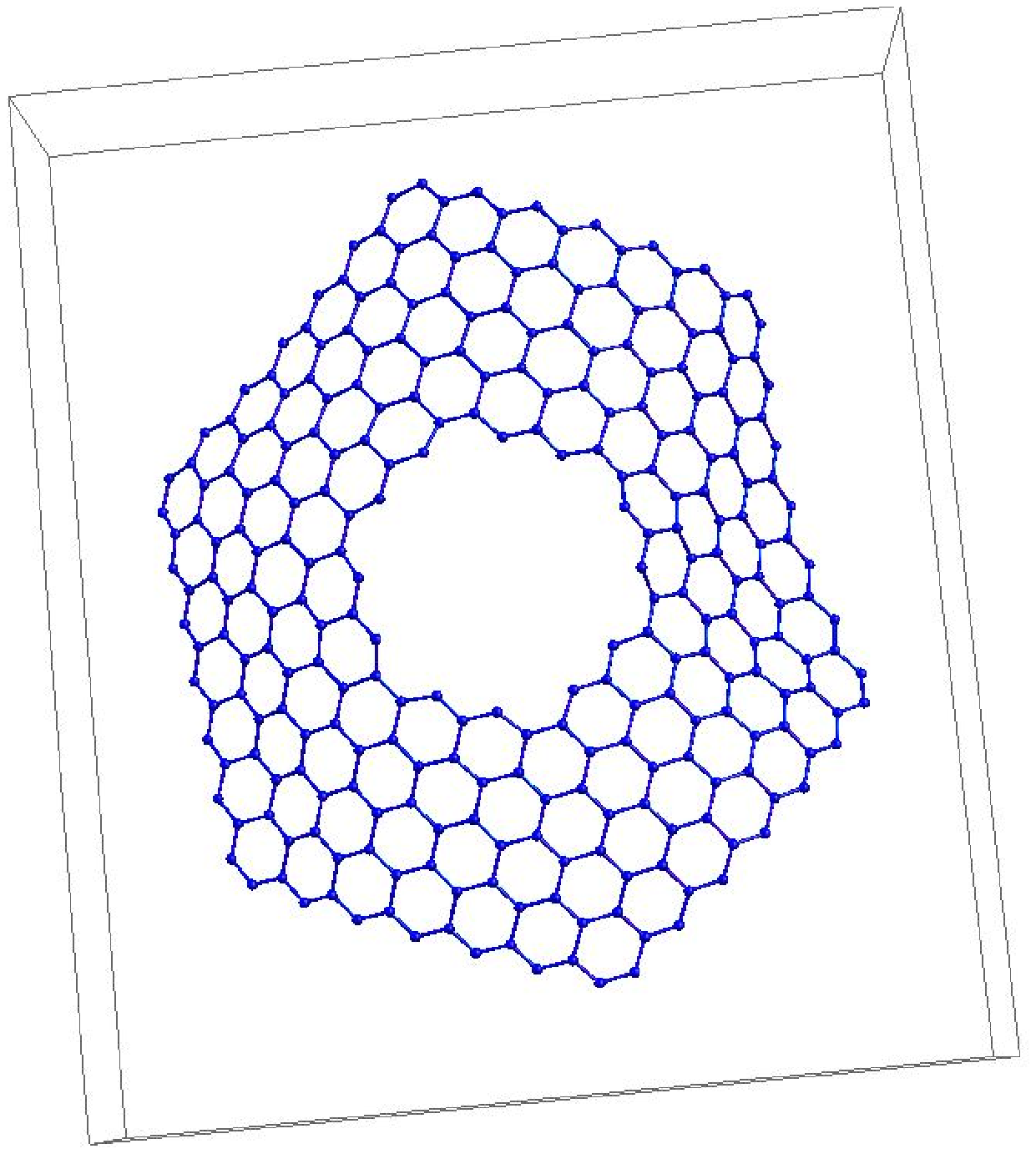}  \hspace{-0.5cm}
\epsfysize 3cm   \epsfbox{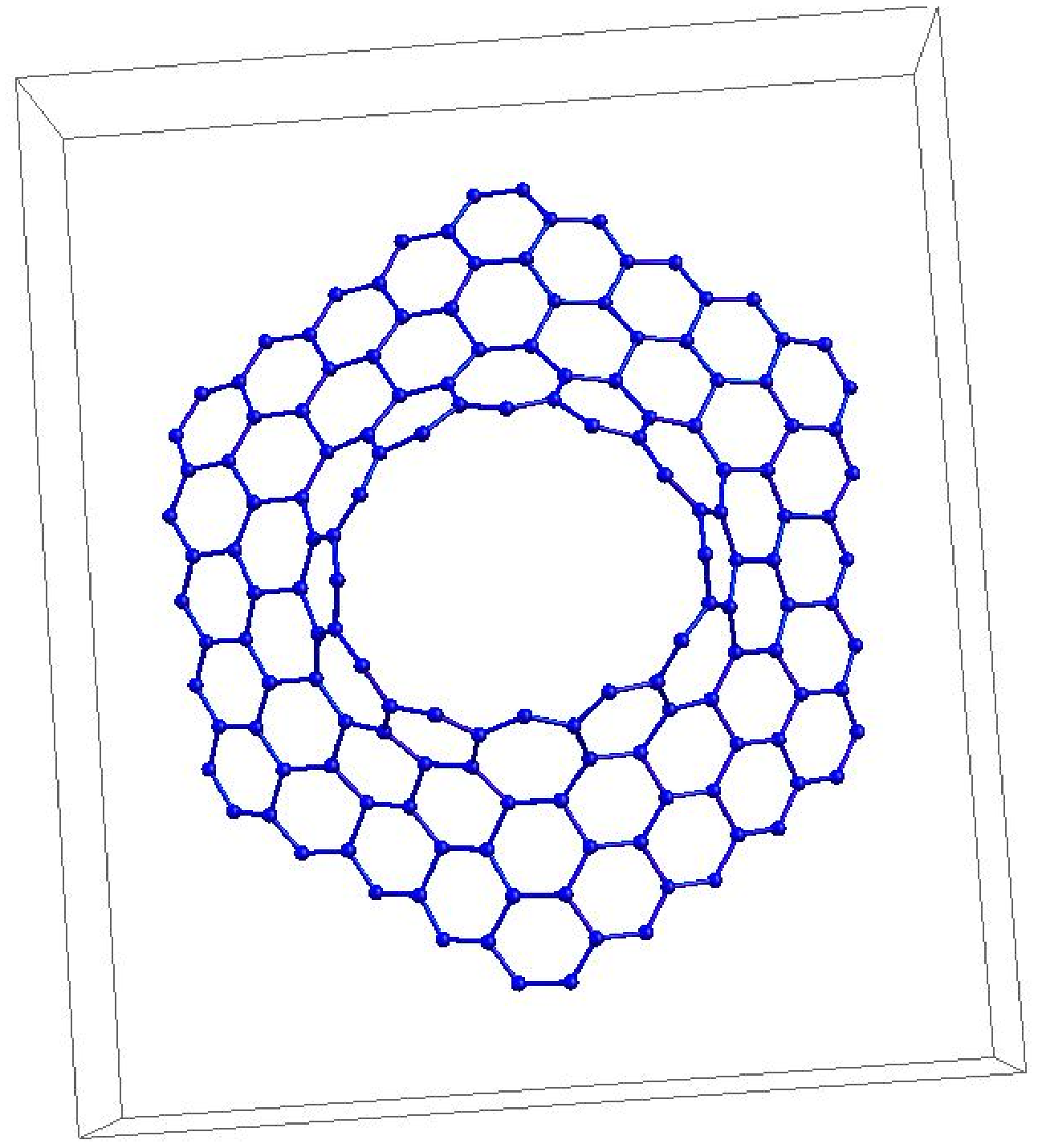}     \hspace{0.5cm} 
 \epsfysize 4cm   \epsfbox{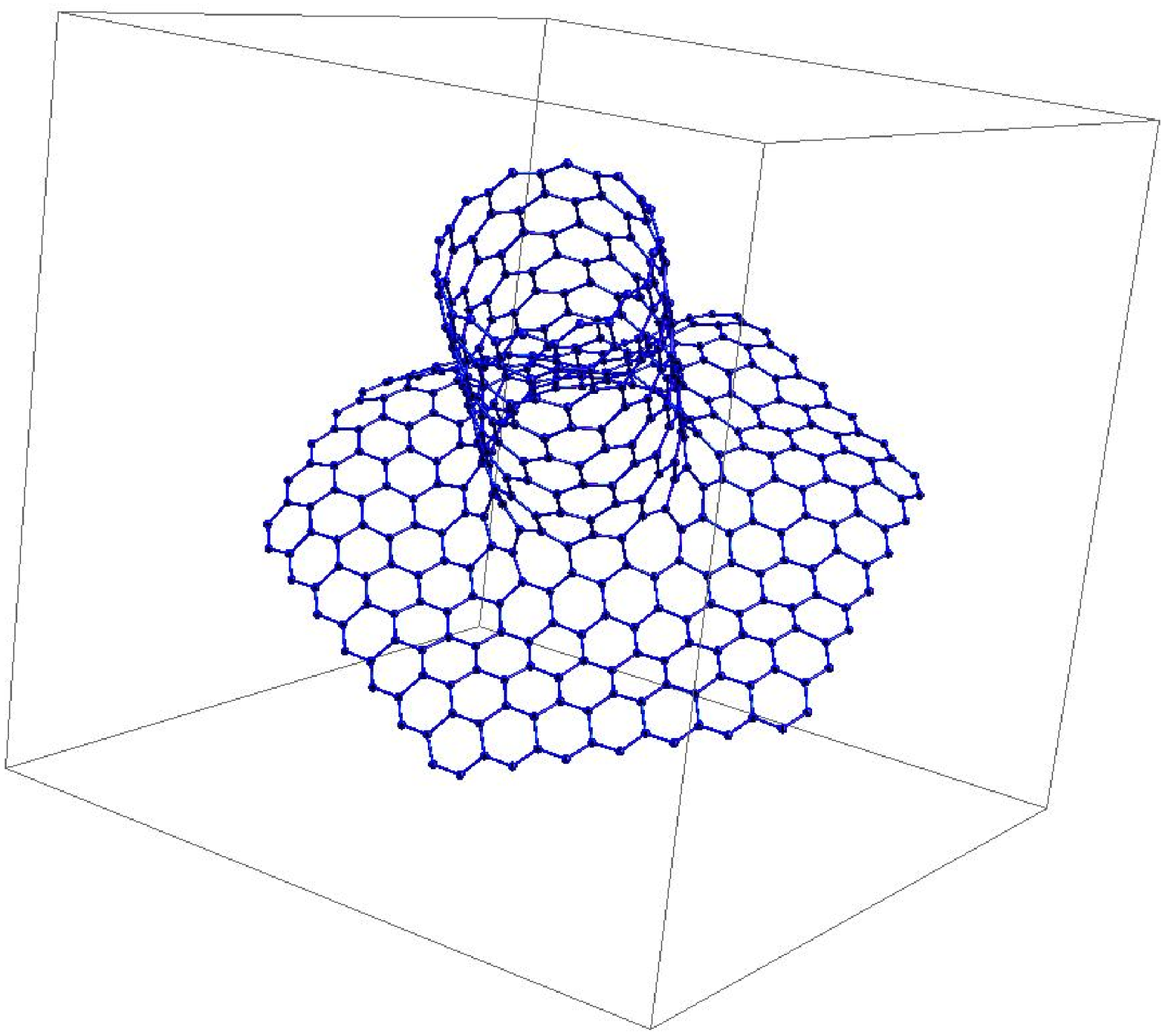}  }   \\
 \hspace{0.65cm}  (a) \hspace{4.75cm} (b)   \hspace{4.75cm} (c)
\end{center}
\caption{Three steps in the construction of a carbon 
nanotube-graphene junction, where it is shown the appearance of the 
6 heptagonal carbon rings at the base of the junction.}
\label{one}
\end{figure}

The simplest way of formally building a nanotube-graphene junction follows the three-step
process represented in Fig. \ref{one}. One can make first a hole with hexagonal shape in
a graphene layer (Fig. \ref{one}(a)). New vertical bonds can be attached then to the 
atoms at the boundary of the hole, to produce next a zig-zag pattern connecting new atoms 
at the other end of the vertical bonds (Fig. \ref{one}(b)). Thus, the lattice is already 
prepared to be extended with a zig-zag nanotube away from the hole (Fig. \ref{one}(c)). 
We see that 6 heptagonal rings are introduced at step 2 of this construction, alternating 
with hexagonal rings at the boundary of the hole. It is clear that the procedure can be 
generalized to attach zig-zag nanotubes of larger radius, while the number of heptagonal 
rings required is always the same. In what follows we will have in mind those graphene
wormholes where the topological defects are regularly distributed along the base of each 
nanotube-graphene junction. The zig-zag nanotubes that can be attached in
this way may have a number of $6n$ hexagons along the waist, corresponding to a $(6n,0)$ 
chirality in the conventional notation to denote the carbon nanotubes\cite{carbon}.

We will take the continuum limit of the graphene wormhole by making the radius 
$R$ of the nanotube much larger than its length. In this way, the wormhole will 
appear at 
large distance scales as the surface of two planes connected at the boundary of a 
common hole. This is the minimal model required to describe the geometry of the 
wormhole, as it disregards possible smooth variations of the curvature in the region 
between the planes. We will see however that such a description already contains 
all the factors needed to reproduce the topological features of the Dirac theory
in the curved space.

In the continuum model of the wormhole geometry, we may start with two coordinate 
charts to cover the two branches (lower and upper plane) of the two-dimensional
manifold. For each of the two charts we may use polar coordinates, that we will 
denote respectively as $r_- , \theta_- $ and $r_+ , \theta_+ $, with the constraint
$r_- , r_+  \geq R $. There is however an obvious way of mapping the upper branch 
with $r_+ \geq R $ into the region corresponding to the hole in the lower branch.
We can make the definition
\begin{equation}
r_- =  \frac{R^2}{r_+}     \;\;\;\;\;  ,  \;\;\;\;   r_+  \geq  R
\label{conf}
\end{equation}
to extend the values of $r_- $ below $R$. In this way, we can pass to describe the 
two-dimensional manifold by means of a single chart, with $0 < r_-  < \infty $. 
The line element $d s$ is given in the region outside the circle $r_- = R$ by the 
conventional metric
\begin{equation}     
ds^2 = dr_-^2 + r_-^2 d\theta_-^2     \;\;\;\;\;   ,  \;\;\;\;   r_-  \geq  R
\end{equation}
while in the inner region it becomes 
\begin{equation}
ds^2 = \left( \frac{R}{r_-} \right)^4
     ( dr_-^2 + r_-^2 d\theta_-^2 )    \;\;\;\;\;   ,  \;\;\;\;   r_-  \leq  R
\end{equation}
It is reassuring to see that the metric thus defined is continuous at $r_- = R$ 
(what would not have been the case if a higher negative power had been used in the 
change of variables (\ref{conf})). The first derivative of the metric tensor is 
however discontinuous, and this is the reflection of the curvature that the 
wormhole concentrates at the junction between the two planes.

From now on we will drop the lower label in the coordinates $r_- , \theta_- $, so 
that we will be using a single radial coordinate with $0 < r  < \infty $. In this
single chart, the metric tensor is given by 
\begin{equation}
g_{\mu \nu } = \Omega^2 (r)  
\left(
\begin{array}{cc}
 1  &  0 \\
 0 &  r^2
\end{array} \right)
\end{equation}  
with the conformal factor
\begin{equation}
\Omega (r) = \left( \frac{R}{r} \right)^2 \theta (R - r) + \theta (r - R)
\end{equation}
From the metric tensor, we obtain the nonvanishing components of the 
connection
\begin{eqnarray}
\Gamma_{r r}^r  & = &  \frac{\Omega '(r)}{\Omega (r)}                         \\
\Gamma_{\theta \theta}^r  & = & - r^2 \frac{\Omega '(r)}{\Omega (r)} - r      \\
\Gamma_{r \theta}^{\theta }  & = &  \frac{\Omega '(r)}{\Omega (r)} + \frac{1}{r}
\end{eqnarray}
One finds that the dependence on $r$ of the above objects is simply proportional to
step functions. Hence, the curvature tensor turns out to be proportional to a 
delta function centered at $r = R$. We get the nonvanishing component      
\begin{eqnarray}
{\cal R}^r_{\theta r \theta }  & = & 
      - r^2  \frac{\Omega ''(r)}{\Omega (r)} 
         + r^2 \left( \frac{\Omega '(r)}{\Omega (r)} \right)^2                       
             - r  \frac{\Omega '(r)}{\Omega (r)}                              \\
                      & = &       - 2 R \;  \delta (r - R)
\end{eqnarray}
From this result, we get finally the scalar curvature 
\begin{equation}
{\cal R} = -  \frac{4}{R} \; \delta (r - R)
\end{equation}

In our minimal model for the wormhole geometry, the curvature is thus localized at 
the contact region between the lower and the upper plane. This is the consequence of 
having taken a continuum limit that compresses the bridge of the wormhole to a circle.
This description is however useful, as it allows us to establish a correspondence 
with the amount of curvature localized at the topological defects in the carbon
lattice. Although the two-dimensional space considered here is infinite, we can still 
compute its Euler characteristic, given the compact support of the scalar curvature. 
Thus, we get
\begin{eqnarray}
\chi   & = &   \frac{1}{4\pi } \int d^2 x  \sqrt{det(g)}  \; {\cal R}            \\
     &  =  &  - \frac{1}{4\pi } \int dr d\theta  \; 4  \delta (r - R)     
       =  - 2
\label{cont}
\end{eqnarray}
We note that the same result is obtained when applying the expression of the 
Euler characteristic for the simplex defined by the carbon lattice of the wormhole. 
The curvature is localized there in the heptagonal carbon rings, in the same way as it 
is so for the pentagonal rings of fullerene molecules\cite{np1}. 
In the case of the fullerenes,
no matter the size, 12 five-fold rings are just needed to curve the carbon lattice into
a spherical cage. As long as the content of curvature of a pentagonal ring is 
the opposite to that of a heptagonal ring, we can assert that a geometry with 12 
heptagonal rings must have an Euler characteristic equal to $-2$, which is the opposite
to that of the sphere. This is consistent with the result (\ref{cont}) obtained for the 
two-dimensional surface, showing that our simplified geometry makes sense as 
a description of the graphene wormhole in the continuum limit.

\section{Gauge fields from topological defects}

In order to set up the correct continuum theory of graphene wormholes, we have
to take into consideration that the effect of the topological defects is not only 
to curve the graphene lattice. This was first realized when looking at the 
low-energy electronic levels of large fullerene molecules. Tight-binding 
calculations carried out in a series of growing clusters showed that a sensible 
continuum limit should account for the existence of two triplets of zero-energy 
modes in the electronic spectrum\cite{full}.
It turned out that such a degeneracy can be actually understood as the effect
of a fictitious gauge field on the curved molecule, arising from the particular 
transformation induced by the disclinations on the electron fields\cite{np1}.

\begin{figure}

\vspace{0.5cm}

\begin{center}
\mbox{\epsfysize 6cm \epsfbox{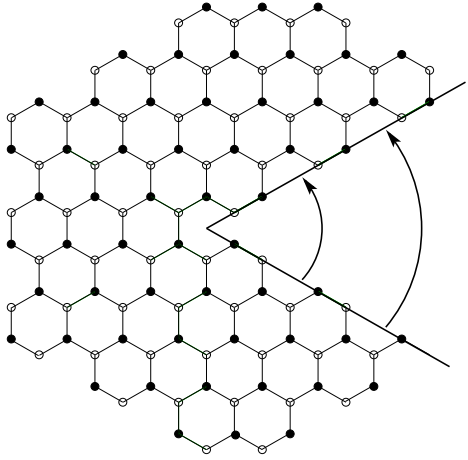}} 

\end{center}
\caption{Formal construction of a pentagonal carbon ring in a 
honeycomb graphene lattice.}
\label{two}
\end{figure}

We start reviewing the nontrivial role of the topological defects by recalling that
the honeycomb graphene lattice has two atoms in the unit cell, in such a way that 
the material is made of two interpenetrating sublattices corresponding to the 
black and white atoms shown in Fig. \ref{two}. Each carbon atom contributes with
one electron to the transport properties, and the tight-binding hamiltonian 
can therefore be written in real space as
\begin{equation}
H = - t \sum_{<i,j>} c_{\bullet i}^{\dagger}c_{\circ j} 
           - t \sum_{<i,j>} c_{\circ i}^{\dagger}c_{\bullet j} 
\label{tb}
\end{equation}
where $c_{\bullet i}^{\dagger} , c_{\circ i}^{\dagger}$ ($c_{\bullet i} , c_{\circ i}$ )
are electron creation (annihilation) operators in the two different sublattices,
and the sum runs over pairs of nearest-neighbor carbon atoms.

When passing to momentum space, the diagonalization of the tight-binding hamiltonian 
leads to the eigenvalue problem
\begin{equation}
     \left(
\begin{array}{cc}
 0  &  -t \sum_{j}\mbox{\Large $e^{i{\bf p \cdot u}_{j}}$} \\
-t \sum_{j}\mbox{\Large $e^{-i{\bf p \cdot u}_{j}}$} &  0
\end{array} \right) \left(
\begin{array}{c}
\psi_{\bullet}  \\
\psi_{\circ}
\end{array}    \right) = \varepsilon ({\bf p})
\left(
\begin{array}{c}
\psi_{\bullet}  \\
\psi_{\circ}
\end{array}    \right)
\label{quad}
\end{equation}
where $\{ {\bf u}_j \}$ is the set of vectors connecting a carbon atom to its 
nearest neighbors. We obtain from (\ref{quad}) the conduction and valence bands 
of graphene
\begin{equation}
\varepsilon ({\bf p}) = \pm t \sqrt{1  
      +  4 \cos^2 \left( \frac{ \sqrt{3}}{2} p_x a \right)
      +  4 \cos \left( \frac{ \sqrt{3}}{2} p_x a \right) 
            \cos \left( \frac{ 3 }{2} p_y a \right)  }
\label{disp}
\end{equation}
where $a$ stands for the C-C distance.

\begin{figure}[h]

\vspace{0.5cm}

\begin{center}
\mbox{\epsfysize 7cm \epsfbox{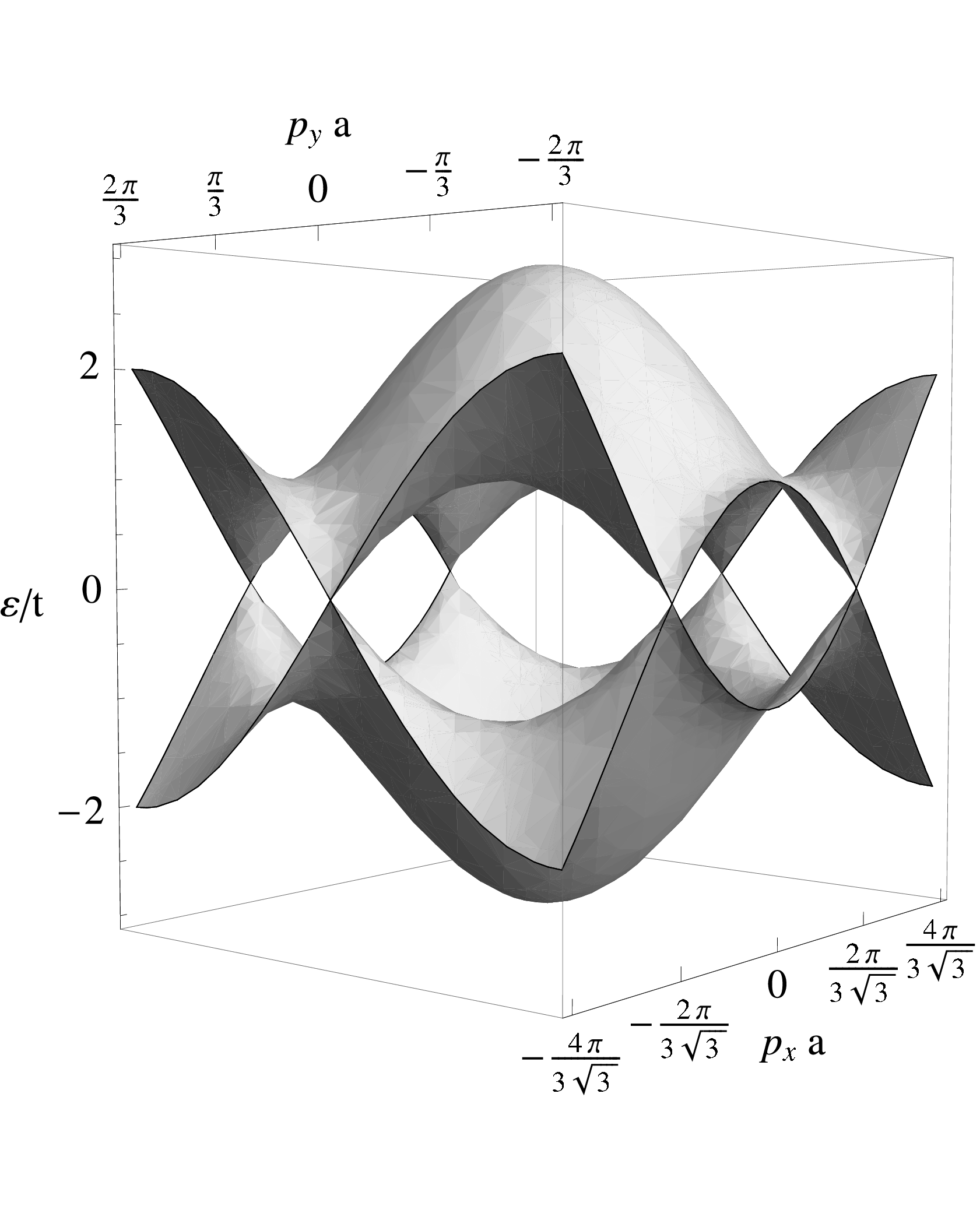}} 

\end{center}
\caption{Plot of the dispersion (\ref{disp}) displaying the conical shape
of the valence and conduction bands at the corners of the hexagonal Brillouin zone.}
\label{three}
\end{figure}

The Fermi level of the undoped system is at $\varepsilon = 0$. The expansion of 
(\ref{disp}) about this level shows that the conduction and valence bands have
conical shape at low energies, meeting at isolated Fermi points 
(so-called Dirac points) which coincide with the corners of the hexagonal Brillouin 
zone (see Fig. \ref{three}). Given the equivalence relation that exists between 
different points in momentum space, it happens that all the low-energy states 
can be accommodated into the conical bands around two inequivalent Dirac points, 
at corners $K$ and $K'$ of the Brillouin zone that can be chosen to make an angle 
of $\pm \pi /3 $ or $\pi $\cite{np1}. This 
implies that the continuum limit of the electron system is given in general by the 
theory of two Dirac fermion fields encoding all the low-energy electronic excitations.

We observe now that the presence of pentagonal or heptagonal rings leads to frustration
in the carbon lattice, in the sense that the atoms cannot be distributed then in two 
different sublattices. We can formally construct a pentagonal defect in the honeycomb
lattice by cutting off a sector of 60 degrees and matching next the two boundaries of 
the remaining space, as represented in Fig. \ref{two}. In this operation, it is clear 
that one is identifying pairs of carbon atoms that belonged to different sublattices
in the original graphene sheet. In the continuum limit, the rotation of 60 degrees 
needed to bring one of the boundaries of the cut graphene plane onto the other can be
properly implemented by introducing a suitable spin connection, realizing the parallel 
transport of the fields when going around the topological defect\cite{np1}. 

There is however an additional effect from the above construction that requires separate 
consideration, as the rotation of 60 degrees leads to a transformation in momentum
space that maps each inequivalent Dirac point into the other. That is, the 
action of matching the two boundaries in Fig. \ref{two} leads to the exchange
of corresponding electron states at the two different Dirac valleys. The imaginary
line that signals the cut in the original graphene plane is just an artifact to define
that operation, as the exchange of the electron states
is only realized when making a complete turn around 
the topological defect. This means that the transformation can be regarded as 
that corresponding to the parallel transport by a gauge field whose flux is concentrated
at the defect\cite{np1}.

A careful consideration of the mapping of the two different Dirac valleys under a rotation 
of 60 degrees shows that the transformation acts on the wavefunctions of the corresponding
electron states at the two Dirac points $K$ and $K'$ as
\begin{equation}
\left( 
\begin{array}{c}
\psi_{K}'   \\
\psi_{K'}'  
\end{array}  \right)
 =  \left(
\begin{array}{cc}
 0  &  1 \\
 -1 &  0
\end{array} \right)
\left( 
\begin{array}{c}
\psi_{K}   \\
\psi_{K'}  
\end{array}  \right)
\label{trans}
\end{equation}
The transformation (\ref{trans}) can be written as the action of a $SU(2)$ 
gauge field ${\bf A}$ on the electrons turning around the topological defect. 
In polar coordinates, the vector potential must have $A_{\theta }$ as its only 
nonvanishing component, fixed by the condition that for any closed path $C$ 
containing the topological defect
\begin{equation}
\mbox{\Large $e^{i\oint_C d\theta A_{\theta }}$} =  \left(
\begin{array}{cc}
 0  &  1 \\
 -1 &  0
\end{array} \right)
\label{rot}
\end{equation} 
The potential ${\bf A}$ is valued on two-dimensional hermitian matrices, and 
it can be viewed as a particular configuration of a non-abelian gauge 
field. We find that
\begin{equation}
A_{\theta} = \frac{\Phi }{2\pi } \tau_2
\label{gauge}
\end{equation}
where $\tau_2 $ is taken from a set of sigma matrices 
$\{ \tau_i   \}$, and $\Phi = \pi /2$.

We observe then that the frustration induced by the pentagonal (as well as 
heptagonal) carbon rings can be mimicked by attaching a line of fictitious
gauge flux $\Phi $ at the topological defect. The corresponding gauge field
acts on the space of the two Dirac points and accounts in general for important
effects in the low-energy part of the electronic spectrum. In particular, 
it plays a crucial role to establish the correct number of zero-energy modes
of the electron system, as shown in the case of the series of growing fullerene
molecules\cite{np1}. 

In what follows, we will need to represent the effect of several heptagonal 
rings on the wormhole geometry. A relevant question is that their combined action
does not lead in general to a gauge field corresponding to the sum of the 
fluxes of the individual defects. This is so as the above representation 
(\ref{rot}) is referred to a particular center of rotation.
The case of two close pentagonal rings has been analyzed in detail in Ref. 
\cite{lc}, making clear that the assessment of their global effect 
requires bringing the action of the respective transformations to a common 
frame. This implies a translation in the honeycomb lattice, that does 
not commute with the action given by (\ref{rot}). The consequence is that
the flux felt away from two topological defects adds to $\pi $ only when the
distance between them is given by a vector $(N,M)$ (in units of the translation
vectors of the honeycomb lattice) such that $N-M$ is a multiple of 3\cite{lc}. 
In the rest of the cases, the total effective flux turns out to be given by 
$\pi /3 $. For our purposes, it will be enough to use this result to compute 
the global effect of the 6 heptagonal carbon rings at each side of the wormhole 
bridge, and to check that their relative location is in correspondence with the
number of zero-energy modes in the spectrum.

\section{Continuum limit of graphene wormholes}

\subsection{Effective field theory}

Our aim is to describe the low-energy regime of the electronic spectrum of 
graphene wormholes, taking the limit in which the radius $R$ of the wormhole 
bridge is much larger than the lattice spacing. The energy levels have to 
correspond then to the eigenvalues of the system of two Dirac fermion fields
on the wormhole geometry, feeling also the fictitious gauge flux attached to 
the topological defects. Regarding this latter effect, we remark that, 
while the vector potential (\ref{gauge}) represents a $SU(2)$ connection,
its action can be made effectively abelian by passing to the basis of 
eigenvectors of $\tau_2 $. We end up in this way with two different
Dirac equations for respective fields $\Psi^+$ and $\Psi^-$, corresponding 
to the two different orientations (outgoing and ingoing) of the fictitious
gauge flux through the surface of the wormhole. We write then the
two eigenvalue problems as
\begin{equation}
i v_F \: \mbox{\boldmath $\sigma $} \cdot (\mbox{\boldmath $\nabla $} \mp
   i {\bf A} )  \Psi^{\pm }   =   \varepsilon  \Psi^{\pm }
\label{dirac}
\end{equation}
$v_F$ being the Fermi velocity and $\pm {\bf A}$ the abelian component of 
the vector potential after projection onto the eigenvectors of 
$\tau_2 $ (we work in units such that $\hbar = 1$).

As long as we are interested in the long-wavelength solutions of (\ref{dirac}),
it will be justified to approximate the effect of the fictitious gauge field by 
its action far away from the wormhole bridge, where the flux coming from the 
topological defects is felt isotropically. Thus, we will assume 
in the resolution of (\ref{dirac}) that the vector potential can be taken 
with a constant nonvanishing component $A_{\theta } = \Phi /2 \pi $. Anyhow,
we know that the effective flux $\Phi $ has to be consistent with the combined
action of the topological defects around the wormhole bridge. According to
what has been stated at the end of the preceding section, such an effective
flux may not correspond in general to the simple addition of individual fluxes
from the defects, but it can be precisely computed from their relative position 
in the wormhole.

As usual for the wave equation of a fermion field, all the vectors in 
(\ref{dirac}) are referred to the tangent space at each point of the two-dimensional
surface. The passage to derivatives with respect to the coordinates $x_{\mu }$
is achieved by making use of the {\em zweibein} $e_a^{\mu }$, defined in terms
of the metric tensor $g_{\mu \nu }$ by
\begin{equation}
g_{\mu \nu } e_a^{\mu } e_b^{\nu } = \delta_{a b }
\end{equation}
The covariant derivative in (\ref{dirac}) can be written as 
$\nabla_{\mu } = \partial_{\mu } + \Gamma_{\mu }$, where the spin connection is
given by\cite{bd} 
\begin{equation}
\Gamma_{\mu } = \frac{1}{8}[\sigma^a ,\sigma^b ] \;  e_a^{\nu } 
     \nabla_{\mu }  e_{b \nu }
\end{equation}
For the particular coordinate system of Sec. II, it is easily found that
\begin{eqnarray}
\Gamma_r           &  =  &   0                                            \\
\Gamma_{\theta }   &  =  &   - \frac{i}{2} \sigma_3 
            \left(  r \frac{\Omega ' (r)}{\Omega (r)} + 1 \right)
\label{spin}
\end{eqnarray}
More explicitly, $\Gamma_{\theta } $ turns out to have a discontinuity at $r = R$, 
from $-(i/2) \sigma_3 $ in the outer region of the circle to 
$(i/2) \sigma_3 $ in the inner region.

Taking into account the form of the spin connection (\ref{spin}) and the gauge
connection (\ref{gauge}), we can split the Dirac equation into two different
expressions corresponding to the lower and the upper sheet of the wormhole:
\begin{equation}
i v_F \left(
\begin{array}{cc}
 0  &  \partial_r - \frac{1}{r} i \partial_{\theta } 
                                  \mp \frac{\Phi }{2\pi r} + \frac{1}{2r}      \\
 \partial_r + \frac{1}{r} i \partial_{\theta } 
                                  \pm \frac{\Phi }{2\pi r} + \frac{1}{2r}   &  0
\end{array} \right)
\left( 
\begin{array}{c}
\Psi_A^{\pm}   \\
\Psi_B^{\pm}  
\end{array}  \right)
= \varepsilon 
\left( 
\begin{array}{c}
\Psi_A^{\pm}   \\
\Psi_B^{\pm}
\end{array}  \right)   \;\;\;\;\;\;\;\;\;\;\;\;  {\rm for}  \;\;\;   r \geq R
\label{dir1}
\end{equation}
and 
\begin{equation}
i v_F \left(  \frac{r}{R}  \right)^2 \left(
\begin{array}{cc}
 0  &  \partial_r - \frac{1}{r} i \partial_{\theta } 
                                  \mp \frac{\Phi' }{2\pi r} - \frac{1}{2r}      \\
 \partial_r + \frac{1}{r} i \partial_{\theta } 
                                  \pm \frac{\Phi' }{2\pi r} - \frac{1}{2r}   &  0
\end{array} \right)
\left( 
\begin{array}{c}
\Psi_A^{\pm}   \\
\Psi_B^{\pm}  
\end{array}  \right)
= \varepsilon 
\left( 
\begin{array}{c}
\Psi_A^{\pm}   \\
\Psi_B^{\pm}
\end{array}  \right)   \;\;\;\;\;\;\;\;\;\;\;\;  {\rm for}  \;\;\;   r \leq R
\label{dir2}
\end{equation}
In the above equations, we can think of $\Psi_A$ and $\Psi_B$ as the 
respective amplitudes of the electron in the two sublattices of the graphene 
lattice, assuming that such a division of the carbon atoms makes sense away 
from the wormhole bridge.

\begin{figure}[h]

\vspace{0.5cm}

\begin{center}
\mbox{\epsfysize 4cm \epsfbox{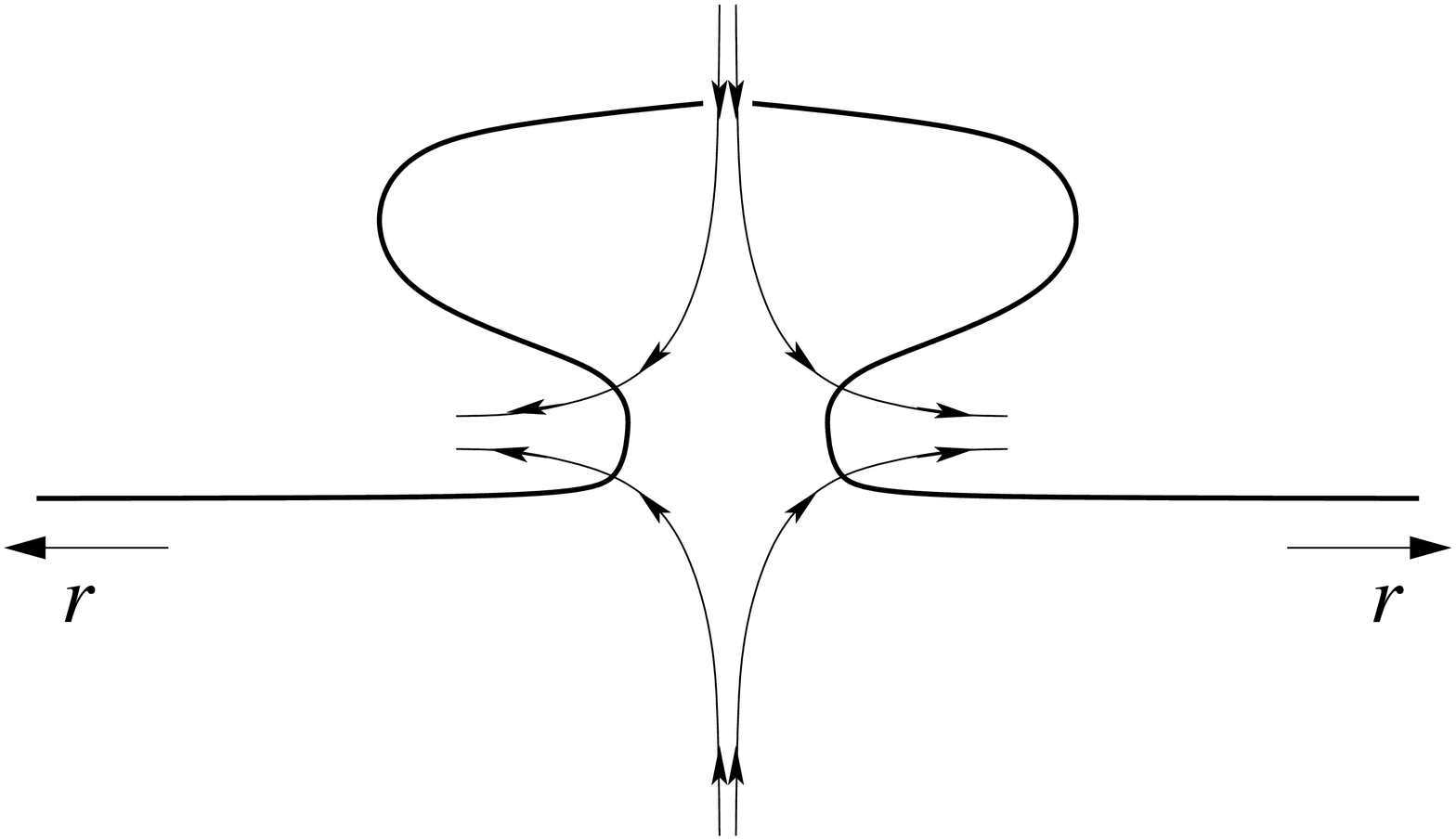}} 

\end{center}
\caption{Schematic representation of the section of a wormhole in the process 
of compactification of the upper sheet into the inner region of the circle 
$r \leq R$. The oriented lines correspond to the fictitious gauge flux 
threading the topological defects.}
\label{four}
\end{figure}

In writing (\ref{dir1}) and (\ref{dir2}), we have introduced different effective
fluxes  $\Phi $ and $\Phi '$ felt by electrons turning respectively in the outer 
and the inner region of the circle $r = R$. The discussion about the actual value of 
these fluxes is rather subtle, as in order to deal with a single coordinate $r$ we
have mapped the upper graphene sheet into the closed region with $r \leq R$. Now 
the electrons turning in this inner region may only feel an effective flux threading 
the origin at $r = 0$, that corresponds to the flux escaping to the point at infinity 
from the original upper sheet. This flux comes from the 6 topological defects at the 
junction with the wormhole bridge, 
as represented schematically in Fig. \ref{four}. This analysis shows that, while 
$\Phi $ has to be effectively the flux induced by the 6 topological defects at the 
junction in the lower graphene sheet, $\Phi '$ must be equal to the opposite flux 
$-\Phi $ threading the origin at $r = 0$.

The preceding argument can be confirmed by reverting the change of variables 
(\ref{conf}) to write Eq. (\ref{dir2}) in the original coordinate $r_+ > R $ of 
the upper graphene branch. According to the above discussion, Eq. (\ref{dir2})
actually reads 
\begin{equation}
i v_F \left(  \frac{r}{R}  \right)^2 \left(
\begin{array}{cc}
 0  &  \partial_r - \frac{1}{r} i \partial_{\theta } 
                                  \pm \frac{\Phi }{2\pi r} - \frac{1}{2r}      \\
 \partial_r + \frac{1}{r} i \partial_{\theta } 
                                  \mp \frac{\Phi }{2\pi r} - \frac{1}{2r}   &  0
\end{array} \right)
\left( 
\begin{array}{c}
\Psi_A^{\pm}   \\
\Psi_B^{\pm}  
\end{array}  \right)
= \varepsilon 
\left( 
\begin{array}{c}
\Psi_A^{\pm}   \\
\Psi_B^{\pm}
\end{array}  \right)   \;\;\;\;\;\;\;  {\rm for}  \;\;\;   r \leq R
\label{dir22}
\end{equation}
By passing to the variable $r_+ = R^2 /r $, we obtain
\begin{equation}
- i v_F \left(
\begin{array}{cc}
 0  &  \partial_{r_+} + \frac{1}{r_+} i \partial_{\theta } 
                                  \mp \frac{\Phi }{2\pi r_+} + \frac{1}{2r_+}      \\
 \partial_{r_+} - \frac{1}{r_+} i \partial_{\theta } 
                                  \pm \frac{\Phi }{2\pi r_+} + \frac{1}{2r_+}   &  0
\end{array} \right)
\left( 
\begin{array}{c}
\Psi_A^{\pm}   \\
\Psi_B^{\pm}  
\end{array}  \right)
= \varepsilon 
\left( 
\begin{array}{c}
\Psi_A^{\pm}   \\
\Psi_B^{\pm}
\end{array}  \right)   \;\;\;\;\;\;  {\rm for}  \;\;\;   r_+ \geq R
\label{upper}
\end{equation}
It is clear that Eq. (\ref{upper}) is formally identical to (\ref{dir1}), up to
a parity transformation accounting 
for the different orientation of the angular variable $\theta $. This simply 
comes from the fact that, when the wormhole is obtained by matching two similar 
graphene branches, the rotation around it is seen as clockwise in 
one of the copies and counter-clockwise in the other. 

In the subsequent discussion, we will need to make a precise assignment for the 
flux $\Phi $ induced by the 6 topological defects at the junction of each graphene 
sheet with the wormhole bridge. As already stated, the total flux is dictated  
by the relative distance $(N,M)$ between heptagonal carbon rings, in such a way
that all graphene wormholes can be classified in two different classes, depending
on whether the value of $N-M$ is a multiple of 3 or not.

\subsection{Zero-energy modes}

We address here the possibility of having bound states 
in the resolution of the Dirac equation (\ref{dirac}). In the case of massive 
Dirac fermions, such localized states correspond to energy levels 
within the gap of the spectrum. As we are dealing with massless fermions, the 
search for bound states has to concentrate on eigenmodes with $\varepsilon = 0$.
On the other hand, the number of these zero-energy modes is in general a measure
of the gauge flux traversing a two-dimensional system. We will see that,
in our case, the existence of the fictitious gauge field provided by the topological 
defects is what gives rise to a certain number of states bound to the wormhole bridge.

In order to determine the effective flux $\Phi $ that appears in Eqs. (\ref{dir1})
and (\ref{dir22}), we have to pay attention to the details of the lattice forming 
the wormhole. As already stated, we will focus here on geometries where the 
topological defects are regularly distributed at both ends of the wormhole bridge. 
When this consists of a small piece of zig-zag nanotube, the construction 
represented in Fig. \ref{one} shows that we can have in general a number $6n$ of 
hexagonal carbon rings around the waist of the tube. This means that the 
relative distance between heptagonal rings will be given then by the couple $(n,0)$, 
following the usual notation in units of the translation vectors of the honeycomb 
lattice\cite{carbon}. 
According to the results of Ref. \cite{lc}, it is only when $n$ is a 
multiple of 3 that the effective flux from the 6 heptagonal rings at each graphene 
branch corresponds to the sum of the individual fluxes. In that instance, we will have 
$\Phi = 3 \pi$. When $n$ is not a multiple of 3, the combined flux of each pair 
of heptagons will only add to $\pi /3 $, giving a total flux $\Phi = \pi $ for the 
6 topological defects at each graphene branch. We will discuss then separately these 
two classes of wormhole geometries in the computation of the zero modes of the Dirac 
equation.

We first consider the case in which the total flux is $\Phi = 3 \pi$. Setting 
$\varepsilon = 0$ in Eqs. (\ref{dir1}) and (\ref{dir22}), we observe that the
eigenmodes get in general a power-law dependence as a function of $r$. Choosing
for instance a solution with $\Psi_A^{\pm} = 0$, we have the equations
\begin{equation} 
\left( \partial_r - \frac{1}{r} i \partial_{\theta } 
                                  \mp \frac{3 }{2 r} + \frac{1}{2r}  \right)
\Psi_B^{\pm}  =  0    \;\;\;\;\;\;\;\;\;\;\;\;  {\rm for}  \;\;\;   r \geq R
\end{equation}
and 
\begin{equation}
\left( \partial_r - \frac{1}{r} i \partial_{\theta } 
                                  \pm \frac{3 }{2 r} - \frac{1}{2r}  \right)
\Psi_B^{\pm}  =  0    \;\;\;\;\;\;\;\;\;\;\;\;  {\rm for}  \;\;\;   r \leq R
\end{equation}

Looking now for solutions that decay at both $r \rightarrow \infty $  and
$r \rightarrow 0$, we find suitable behaviors for $\Psi_B^{-}$
\begin{eqnarray}
 \Psi_B^{-}  & \sim &  r^{-l-2} e^{il \theta }        \;\;  ,  \;\;   r \geq R    \\
 \Psi_B^{-}  & \sim &  r^{-l+2} e^{il \theta }        \;\;  ,  \;\;   r \leq R 
\end{eqnarray}
It is clear then that these modes lead to states localized at the wormhole
for $l = 0$ and $\pm 1$. A more detailed inspection shows however that only the 
state with $l = 0$ is strictly normalizable. In the curved space, the finiteness
of the norm of the eigenstates is enforced by the condition
\begin{equation}
\int d^2 x \sqrt{det(g)}  \left| \Psi ({\bf r}) \right|^2   < \infty 
\end{equation}
In our case, taking into account the explicit form of the determinant of the 
metric tensor leads to the constraints for each angular momentum $l$
\begin{eqnarray}
\int_{R}^{\infty }  dr \; r \; (r^2)^{-l-2}  & < &  \infty    \label{n1}   \\
\int_0^{R}  dr \; \frac{1}{r^3} \; (r^2)^{-l+2}        & < &  \infty
\label{n2}
\end{eqnarray}
We observe that the modes with $l = \pm 1$ cannot fulfil both (\ref{n1}) and
(\ref{n2}). Their norm will only show however a mild logarithmic divergence as 
a function of the size of large but finite systems. This will still make 
possible to find a signature of these modes, when looking for localized states 
in the numerical study of large graphene wormholes. 

The above discussion applies to the particular case of taking 
$\Psi_A^{\pm} = 0$. It is clear however that similar solutions
can be obtained by replacing the lower by the upper component of
the Dirac spinor, while inverting the direction of the fictitious flux and
the angular momentum number, as these combined operations represent an exact 
symmetry of the Dirac equation. We see therefore that the above localized 
states have their respective partners with $\Psi_A^{+} \neq 0$, corresponding 
again to angular momentum $l = 0$ and $\pm 1$.

A similar analysis can be carried out for wormhole geometries with $(6n,0)$
nanotubes in which $n$ is not a multiple of 3. In these cases, the effective 
flux from the topological defects is reduced down to $\Phi = \pi $. Taking
first $\Psi_A^{\pm} = 0$, the Dirac equation for the zero-energy modes
becomes now 
\begin{equation} 
\left( \partial_r - \frac{1}{r} i \partial_{\theta } 
                                  \mp \frac{1 }{2 r} + \frac{1}{2r}  \right)
\Psi_B^{\pm}  =  0    \;\;\;\;\;\;\;\;\;\;\;\;  {\rm for}  \;\;\;   r \geq R
\end{equation}
and 
\begin{equation}
\left( \partial_r - \frac{1}{r} i \partial_{\theta } 
                                  \pm \frac{1 }{2 r} - \frac{1}{2r}  \right)
\Psi_B^{\pm}  =  0    \;\;\;\;\;\;\;\;\;\;\;\;  {\rm for}  \;\;\;   r \leq R
\end{equation}

We find now more tight conditions to ensure the decay of the eigenmodes 
at $r \rightarrow \infty $ and $r \rightarrow 0$, as the resolution of the above
equations gives the behaviors 
\begin{eqnarray}
 \Psi_B^{-}  & \sim &  r^{-l-1} e^{il \theta }        \;\;  ,  \;\;   r \geq R    \\
 \Psi_B^{-}  & \sim &  r^{-l+1} e^{il \theta }        \;\;  ,  \;\;   r \leq R 
\end{eqnarray}
It turns out that, in this case, a localized state exists only for $l = 0$.
Its norm is not convergent in the infinite wormhole geometry but, 
as also happened in the preceding discussion, it just shows a soft logarithmic 
divergence that can make the mode to appear as a localized state in finite 
real systems. We complete this analysis by remarking that a similar
mode arises, now with $\Psi_A^{+} \neq 0$, when the orientation of the 
fictitious gauge flux is inverted in the Dirac equation. 

The above search of the zero-energy modes shows that the number of localized states 
of the Dirac equation bears a direct relation to the gauge flux threading the 
two-dimensional system. This property is at the basis of the development of the 
Landau levels in a conventional transverse magnetic field, and it is confirmed 
in our system with fictitious gauge flux attached to the topological
defects. The number of zero modes is actually a topological invariant, 
not sensitive to the particular details of the flux over the two-dimensional
geometry. This is why their signatures can be found in the real carbon lattices
where the effect of the fictitious gauge flux is not felt isotropically, 
as we will see in the numerical study of the graphene wormholes.

\section{Numerical approach to graphene wormhole spectra}

We want to check that the above features derived from the Dirac equation
in the curved space are actually present in real graphene wormholes. For 
this purpose, we will adopt a numerical tight-binding approach to describe
the electronic structure of the curved carbon material. Thus, we will assume
that the electronic properties can be modeled by the 
hopping of electrons between nearest-neighbor atoms of the carbon lattice,
following the same approach leading to Eq. (\ref{tb}). The tight-binding
hamiltonian reads now
\begin{equation}
H = - t \sum_{<i,j>} c_{i}^{\dagger}c_{j} 
\label{tbc}
\end{equation}
where $c_{i}^{\dagger}$ ($c_{i}$) are electron creation (annihilation) operators
and the sum runs over pairs of nearest-neighbor carbon atoms in the graphene
wormhole. The lattices we 
are going to consider are obtained by matching two nanotube-graphene junctions 
of the type shown in Fig. \ref{one}(c). This means that we will always have 
a short zig-zag nanotube bridge connecting the two graphene branches of the 
wormhole. We will take however this nanotube with a length much smaller
than its radius, so that a sensible comparison can be achieved with the 
previous analytic approach to the wormhole in the continuum limit.   
  
We have carried out the numerical diagonalization of the hamiltonian 
(\ref{tbc}) in lattices containing of the order of $\sim 100,000$ atoms, for 
different values of the wormhole radius. In practice, we have matched the
two nanotube-graphene junctions of the wormhole by aligning pairwise the position 
of heptagonal carbon rings at the two ends of the nanotube bridge. This choice
has the advantage of preserving in the wormhole the symmetry of the nanotube-graphene
junctions under rotations around the nanotube axis. In this way, it is possible
to classify the different energy levels according to their eigenvalue $q$ 
under a rotation of $\pi /3 $. The spectrum can be decomposed then in several 
sectors corresponding to the different values $q = 1, e^{\pm i \pi /3},
e^{\pm 2i \pi /3}$ and $-1$. This turns out to be very convenient for the purpose of
drawing the connection with the eigenmodes having well-defined angular momentum
in the continuum limit. 

The inspection of spectra of different graphene wormholes shows that they fall
into two broad classes, paying attention to the electronic structure in the 
different sectors labeled by the eigenvalue $q$. In this respect, a 
convenient way of characterizing the electronic system is to look at the 
features produced by the wormhole bridge, as this is the source of new
electronic properties. For each wormhole, we have then computed preferentially 
the local density of states 
for a circular ring of atoms around one of the ends
of the nanotube bridge. This has been complemented with the analysis of the 
spatial decay of the eigenmodes in the graphene sheets, in order to discern
the propagating or localized character of the electronic states.

\begin{figure}
\begin{center}
\mbox{
\epsfxsize 4cm \epsfbox{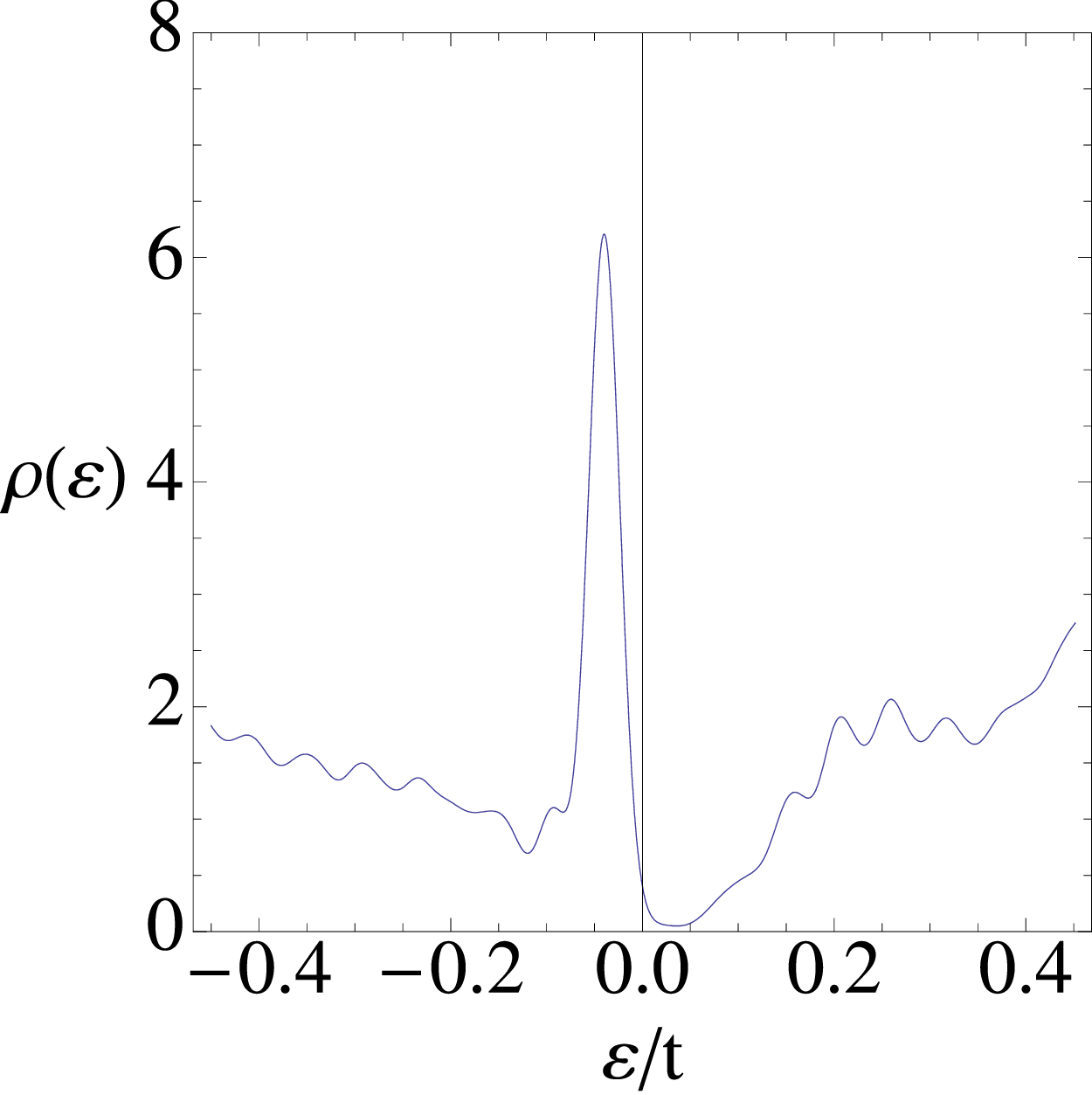}  \hspace{0.5cm}
\epsfxsize 4cm   \epsfbox{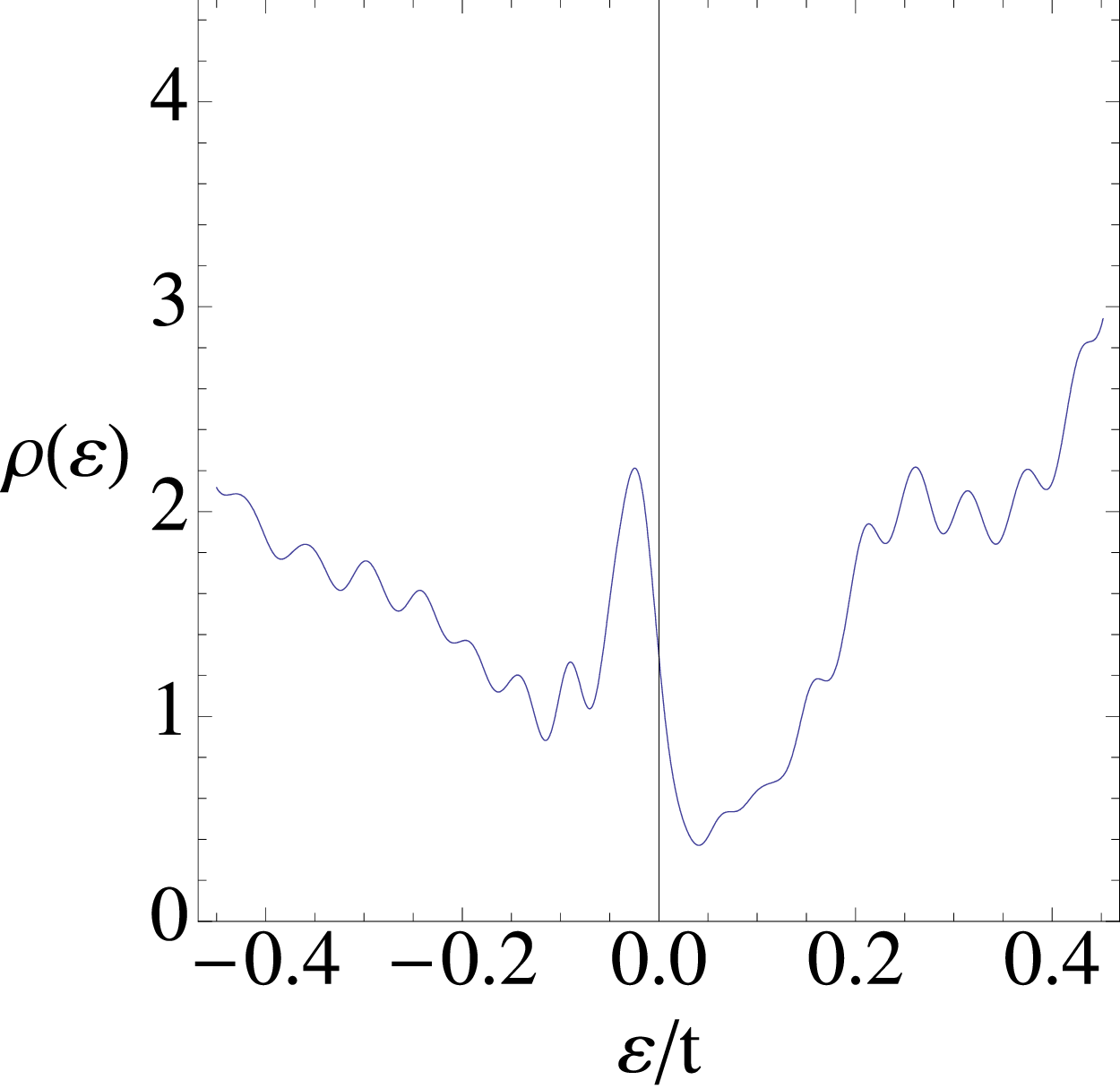}  }\\
 \hspace{0.65cm}  (a) \hspace{4.75cm} (b)   \\  \mbox{}  \\
\mbox{
\epsfxsize 4cm \epsfbox{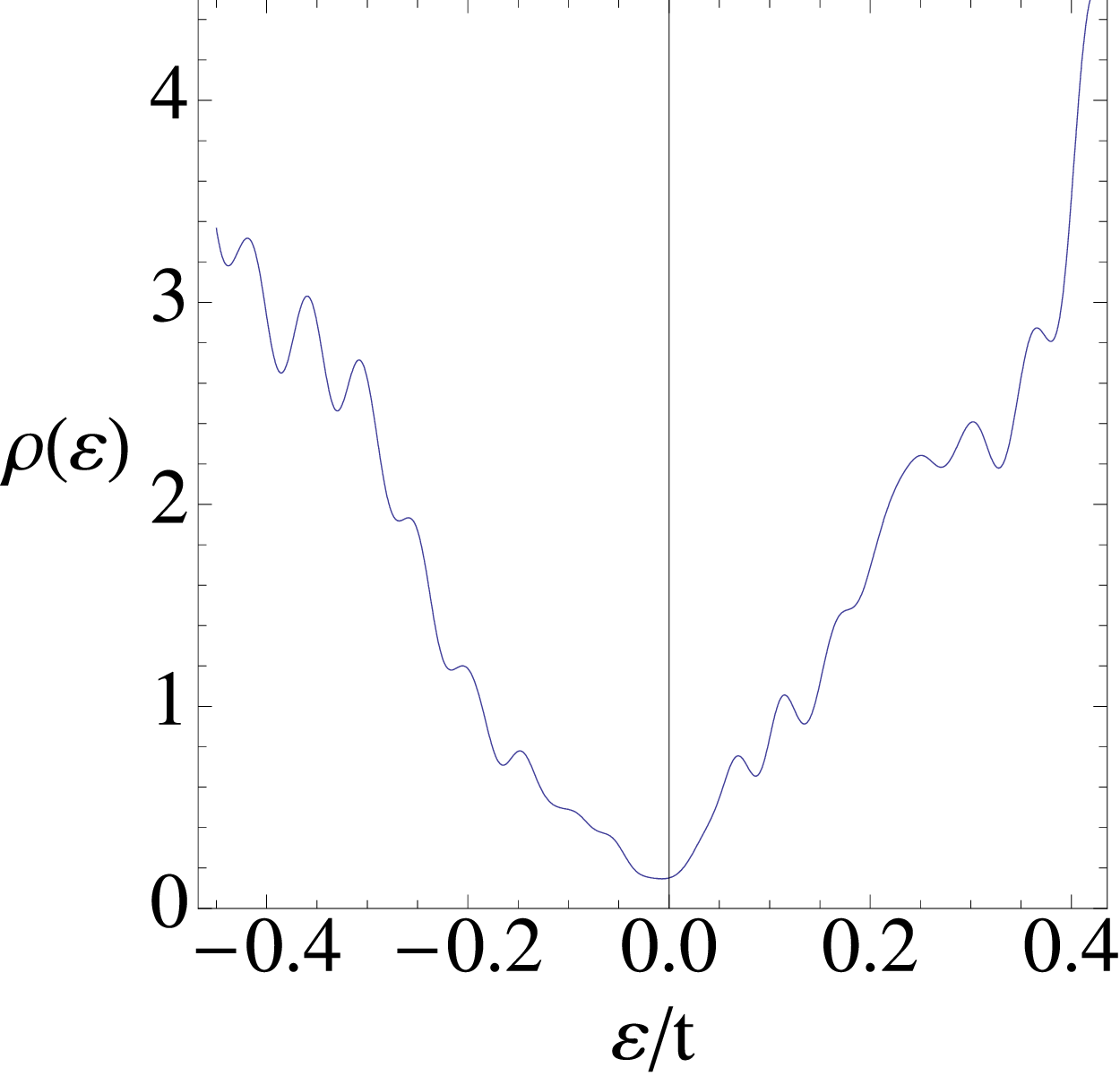}  \hspace{0.5cm}
\epsfxsize 4cm     \epsfbox{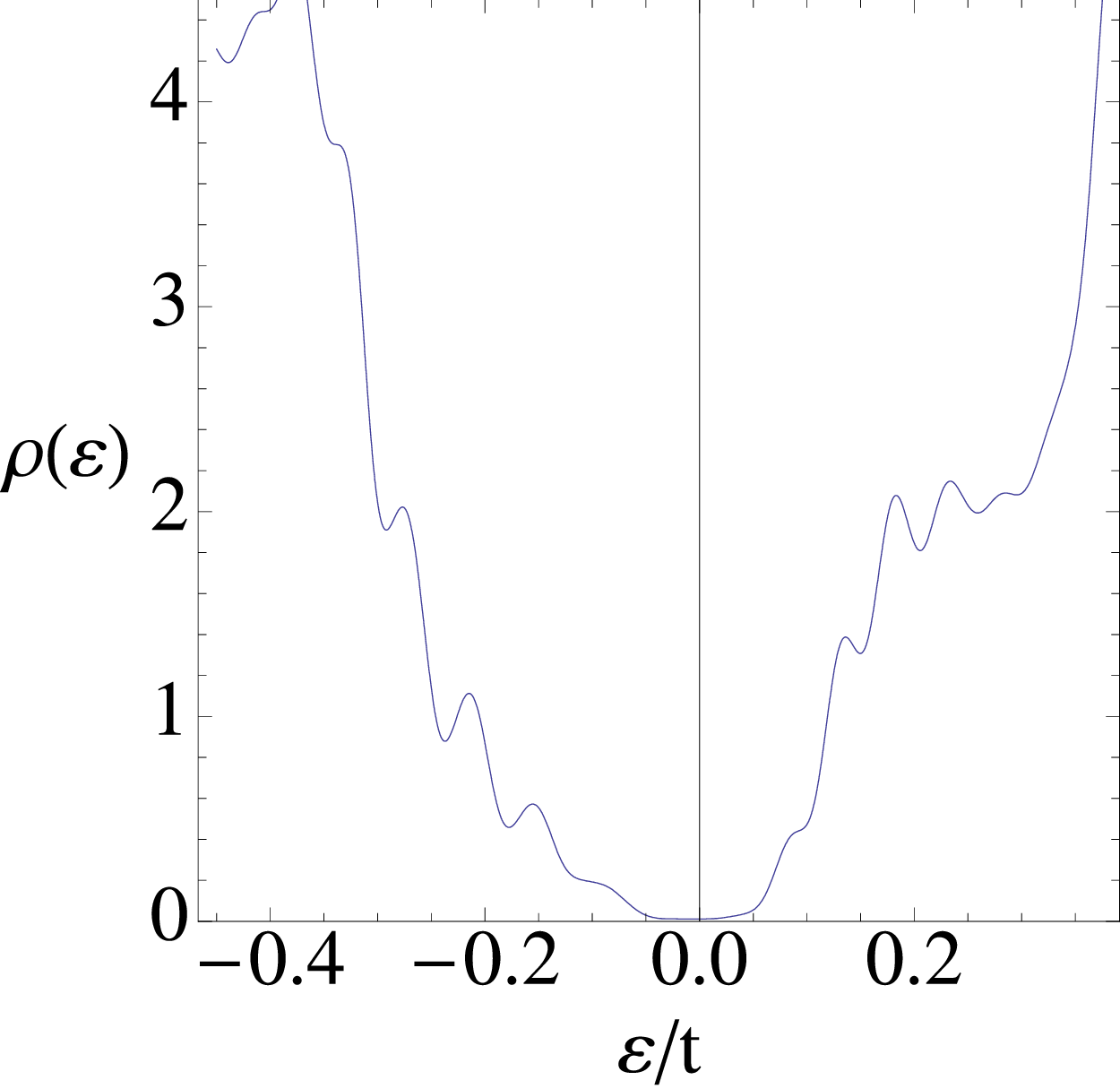} }\\
 \hspace{0.65cm}  (c) \hspace{4.75cm} (d)
\end{center}
\caption{Sequence of local densities of states for a circular ring of atoms 
around one of the ends of a wormhole bridge made of a short zig-zag (54,0) 
nanotube, for the different sectors corresponding to eigenvalue $q$ under
$\pi /3$ rotation equal to 1 (a), $e^{\pm i\pi /3}$ (b),
$e^{\pm 2i\pi /3}$ (c), and $-1$ (d). Energy is measured in units of the
transfer integral $t$.}
\label{five}
\end{figure}

We show for instance in Fig. \ref{five} the mentioned local density of states
$\rho (\varepsilon )$
for a graphene wormhole with a short zig-zag $(54,0)$ nanotube bridge. The 
length of the bridge amounts to two nanotube unit cells in this case, complying 
therefore with the requirement of being much smaller than the radius of the 
wormhole. We observe that the density of states is in general suppressed around 
$\varepsilon = 0$, while there are prominent peaks at very low energy in 
the sectors corresponding to $q = 1$ and $e^{\pm i \pi /3 }$. The depletion in
the density of states is consistent with the fact that the underlying effective
theory in the continuum limit must be that of a pair of Dirac fermion fields, 
which have in general a spectrum with a vanishing density of states at the 
Dirac point. Furthermore, 
we have to bear in mind that, in the representations of the density of states, 
the contribution of each level has been smoothed with a gaussian broadening in 
order to give a continuous appearance to the plots. Thus, a closer look reveals 
that a couple of almost degenerated levels is responsible for the peak in the 
sector $q = 1$, while that for $q = e^{i \pi /3 }$ (or $q = e^{-i \pi /3 }$) 
comes essentially from the contribution of two other levels with very low energy.  
Further analysis shows that the corresponding spatial distribution for these 
modes has a maximum at the wormhole bridge, with a pronounced decay 
when moving away along the radial direction in each graphene sheet.

We have checked that the peaks in the local density of states for $q = 1$ and 
$e^{\pm i \pi /3 }$ are always present when the graphene wormholes have
a $(6n, 0)$ nanotube bridge such that $n$ is a multiple of 3. Quite remarkably,
these are the instances in which the fictitious gauge flux from the heptagonal
carbon rings reaches its maximum value, leading to two triplets of zero-energy
modes with angular momenta $l = 0$ and $\pm 1$ in the effective continuum theory. 
This is then in agreement with the observation of the localized states
in the numerical diagonalization of the graphene wormholes. We recall 
that, in the effective theory, the angular momentum $l$ gives the smooth spatial
modulation corresponding to the envelope of the electron wavefunction. 
This has in general fast oscillations dictated by the large momenta 
$p_x = \pm 4 \pi /3 \sqrt{3} a$ at the Dirac points\cite{np1}. In the case 
of zig-zag $(6n, 0)$ nanotubes with $n$ equal to a multiple of 3, however, the 
oscillations in the lowest-energy branches of the spectrum turn out to correspond 
to the eigenvalue $q = 1$. This explains why the localized states with $l = 0$ 
and $\pm 1$ appear in the
lattice for such a value of $q$ and those obtained by adding or subtracting 
one unit of angular momentum.

\begin{figure}
\begin{center}
\mbox{
\epsfxsize 4cm \epsfbox{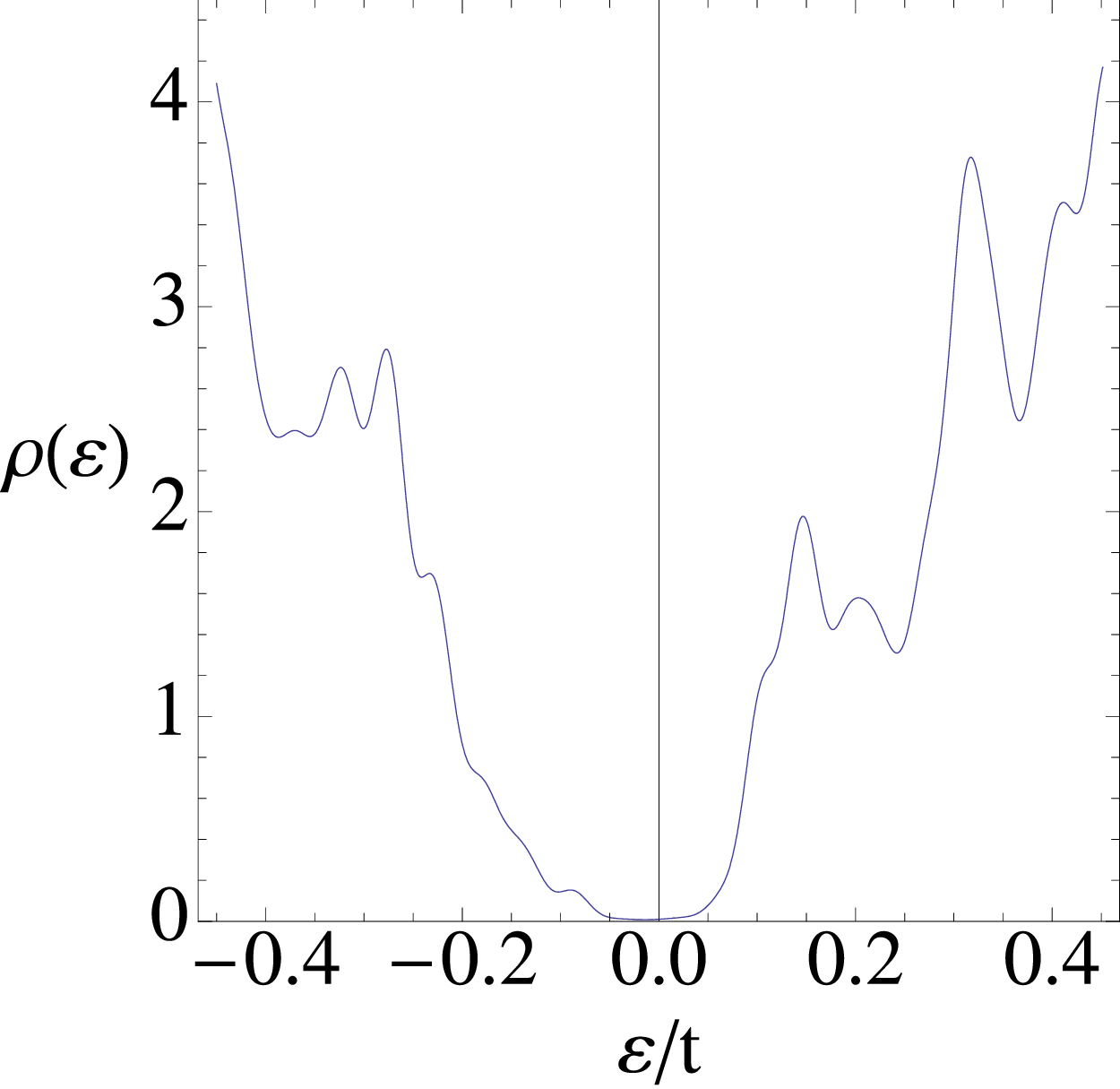}   \hspace{0.5cm}
\epsfxsize 4cm   \epsfbox{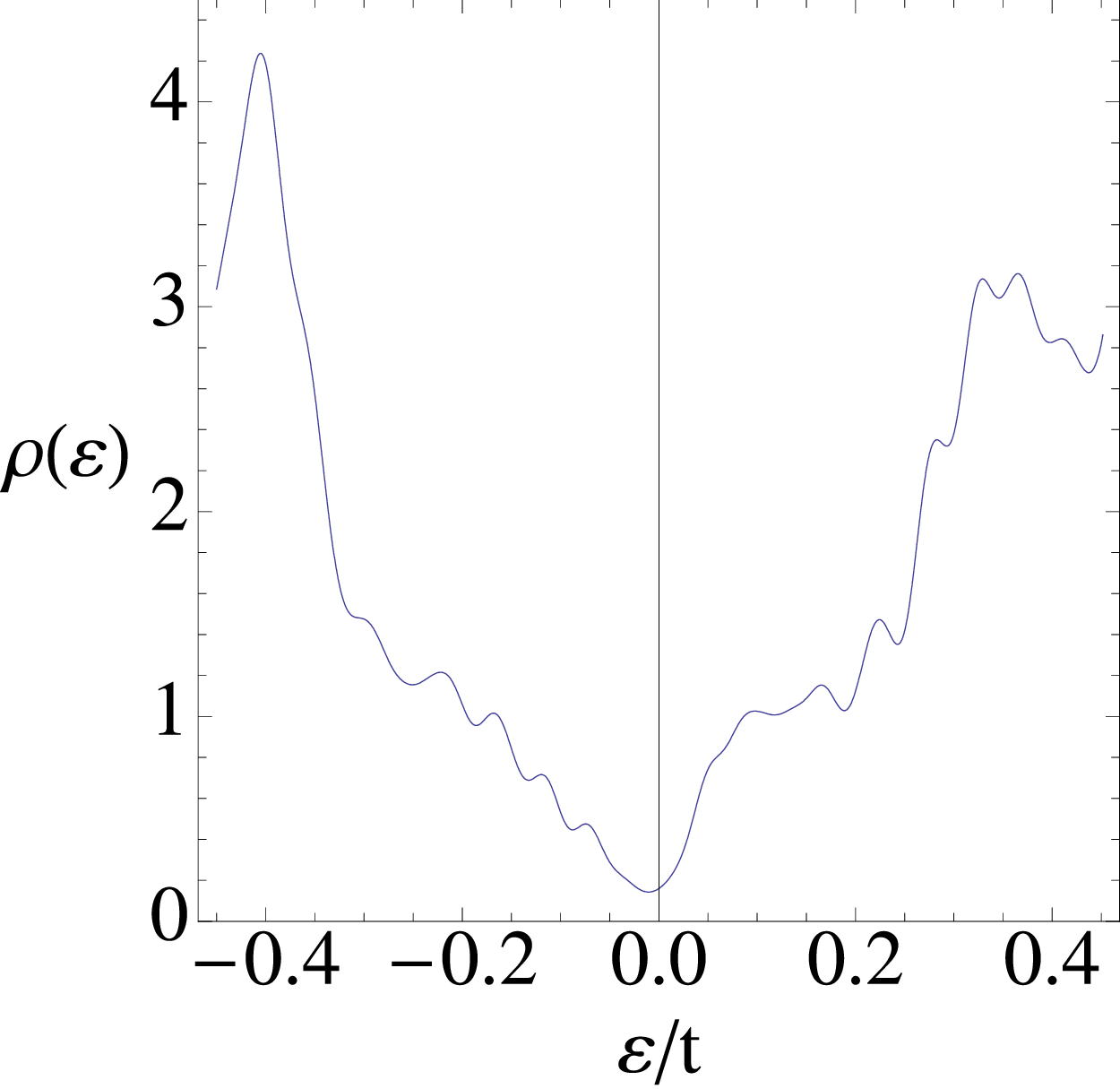}  }\\
 \hspace{0.65cm}  (a) \hspace{4.75cm} (b)   \\   \mbox{}  \\
\mbox{
\epsfxsize 4cm \epsfbox{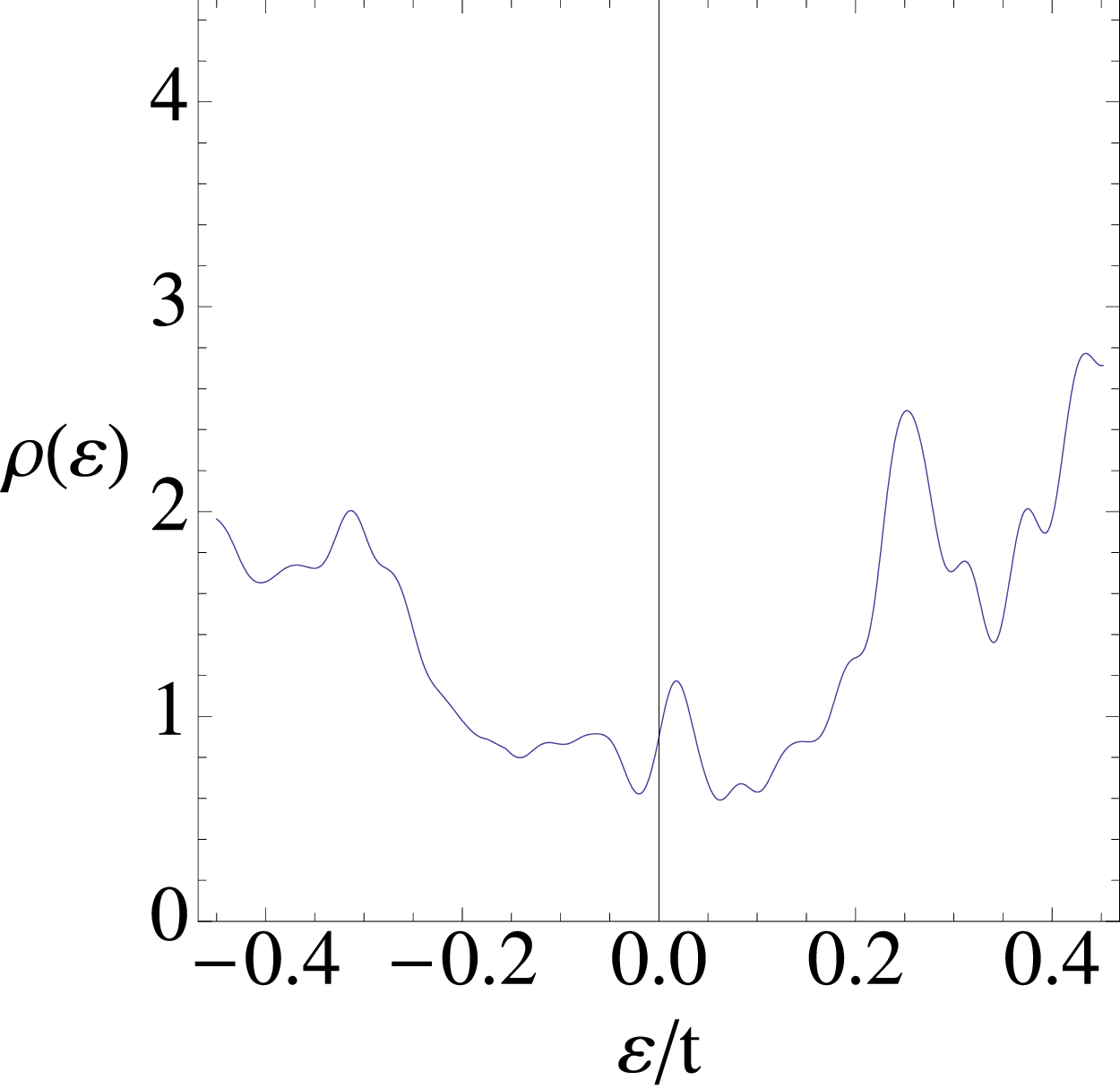}   \hspace{0.5cm}
\epsfxsize 4cm    \epsfbox{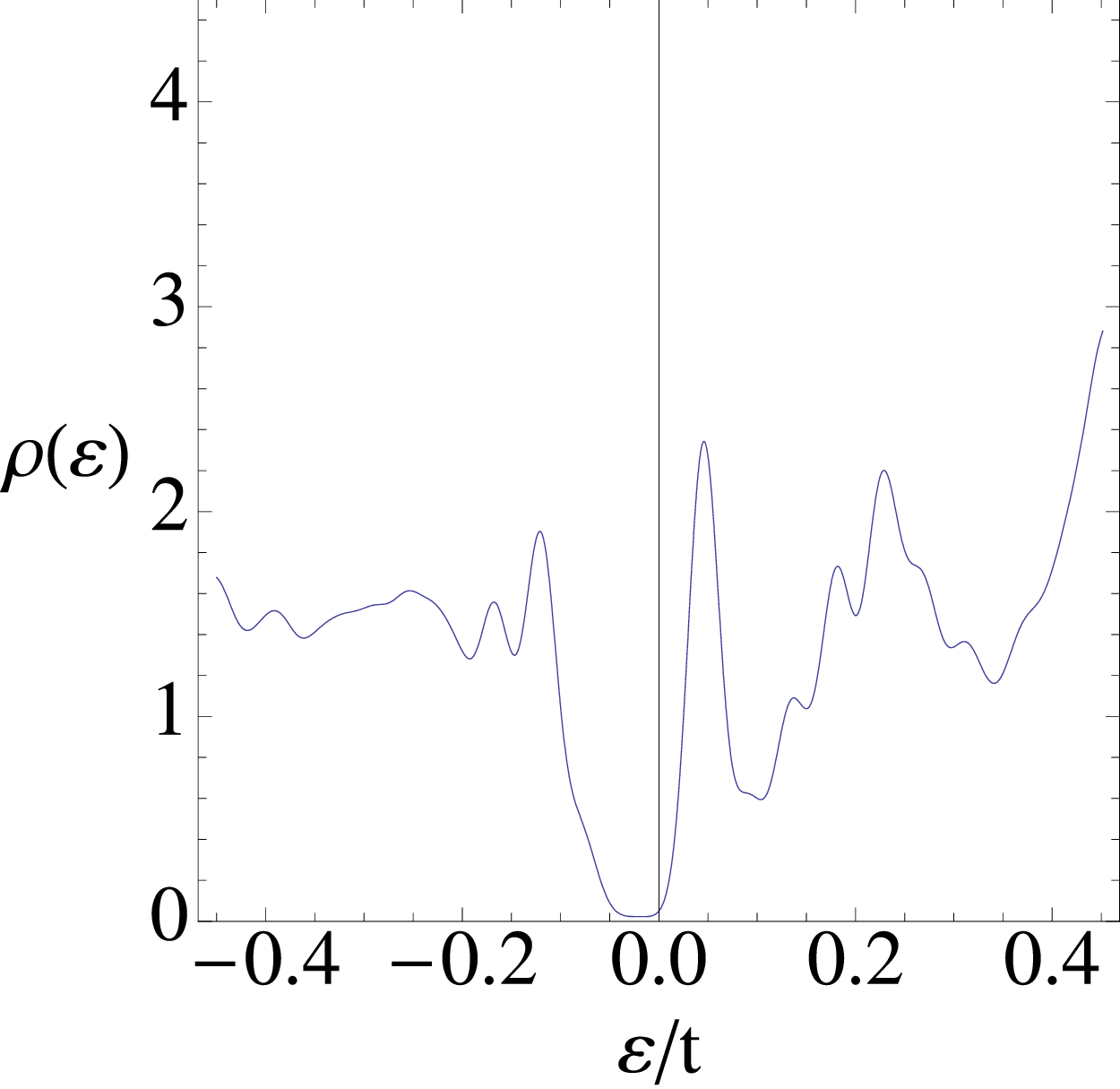} }\\
 \hspace{0.65cm}  (c) \hspace{4.75cm} (d)
\end{center}
\caption{Similar sequence as in Fig. \ref{five} for a wormhole bridge made
of a short zig-zag (48,0) nanotube, for $q = 1$ (a), $e^{\pm i\pi /3}$ (b),
$e^{\pm 2i\pi /3}$ (c), and $-1$ (d).}
\label{six}
\end{figure}

The rest of graphene wormholes, with a zig-zag $(6n,0)$ nanotube bridge 
where $n$ is not a multiple of 3, display spectra with a different but 
distinctive behavior. This is illustrated in Fig. \ref{six} in the case of a
wormhole made of a short $(48,0)$ nanotube. The local density of states 
$\rho (\varepsilon )$ is again depleted
in general around $\varepsilon = 0$, but it does not vanish now in the sectors with 
$q = e^{\pm 2i\pi /3 }$, showing even a small peak at very low energy. This can be 
explained by the fact that, in the $(6n,0)$ nanotubes with $n$ not equal to a 
multiple of 3, the lowest-energy branches have in general large angular momentum 
numbers that do not correspond to the eigenvalue $q = 1$. In the zig-zag nanotubes,
the electrons pick up a large transverse momentum $p_{\perp } = 4 \pi /3 \sqrt{3} a$ 
at the Dirac point, which translates when turning around the nanotube waist of length 
$6 \sqrt{3} na$ into a variation of the phase $\Delta \phi = 8 \pi n$ 
for the electron wavefunction. If $n$
is not a multiple of 3, the symmetry of these lowest-energy states under a rotation 
of $\pi / 3$ makes them to fall always in the sector with $q = e^{\pm 2i\pi /3 }$.  
It seems therefore that, though the nanotube bridge has a short length in our 
lattices (extending just four nanotube unit cells in the example of Fig. \ref{six}), 
it plays a crucial role dictating the symmetry of the low-energy states in the 
wormhole.

It can be checked that the small peak in the sector with $q = e^{ 2i\pi /3 }$
in Fig. \ref{six} (and its counterpart for $q = e^{- 2i\pi /3 }$) 
arises from the presence of a state with very low energy localized at the wormhole.
This agrees with the prediction of the effective continuum theory developed 
above, bearing in mind that the states with $l = 0$ have their spatial oscillations
dictated by the large momentum at the Dirac point, which leads in this case to 
$q = e^{\pm 2i\pi /3 }$.
The smallness of the peak is partly due to the fact that it comes from the 
contribution of a single level for the given value of $q$, but it can be also taken
as a reflection that the localized mode cannot give rise to a truly normalizable 
state in the continuum theory. 
On the other hand, we remark that other low-energy features like the innermost 
peaks at both sides of the Fermi level in Fig. \ref{six}(d) have to do instead with
the resonant transmission through the nanotube bridge. This is consistent with the
fact that similar peaks have been already observed in that sector of the local density
of states in carbon nanotube-graphene junctions\cite{ngj}. The presence of the 
nanotube bridge is in general responsible for those sharp enhancements in the 
local density of states, which mark the thresholds at which new channels open 
for transmission.

Going back to Fig. \ref{five}, we observe that the much larger peak found in the sector 
with $q = 1$, in comparison to those for $q = e^{\pm  i \pi /3 }$ in that figure, 
is actually the signature of the strong localization of a couple of states in the 
graphene wormhole, in agreement with the prediction of two well-defined bound 
states in the continuum theory. This further supports the correspondence between the 
low-energy features observed in the numerical diagonalization of the tight-binding 
hamiltonian and the results from the effective Dirac theory, confirming the validity 
of the latter for the continuum limit description of the curved carbon lattices.

\section{Conclusions}

In this paper we have developed an effective field theory accounting for the
continuum limit of graphene wormholes, when their radius is much larger than the
length of the nanotube bridge. Under this condition, we have dealt with a minimal 
continuum model of 
the wormhole geometry obtained by matching two two-dimensional sheets at the boundary 
of a common circular hole. This provides a sensible description of the properties 
of the lattice geometry when looking over large length scales,
as the graphene wormholes have their intrinsic curvature concentrated on 12 
heptagonal carbon rings at the junctions of the graphene branches with the nanotube
bridge. A nice check of the consistency of the continuum model is made by realizing
that the integral of the curvature over the two-dimensional space is an integer
number (in units of $4 \pi $) matching the value expected from the sum of the
contributions of the heptagonal carbon rings in the lattice geometry.

We have seen that the continuum limit has to be complemented by
including the effect that the heptagonal carbon rings induce on the Dirac fields
encoding the low-energy electronic excitations of the carbon material. This action 
can be mimicked by attaching a line of fictitious gauge flux at each topological
defect, as already illustrated in the case of the fullerene lattices\cite{np1}. 
The graphene
wormholes represent actually an instance which is to some extent dual to the case of 
the fullerenes, as the 12 pentagonal carbon rings in those closed lattices play a role 
opposite to that of the heptagonal defects in the wormhole. It is remarkable that 
the zero-energy modes of the Dirac equation in the continuum model of the wormhole 
can be arranged into two triplets, when the fictitious gauge flux traversing the 
two-dimensional space becomes maximal (and actually equal to the radial flux 
supported by the fullerene molecules). We can think of those low-lying levels 
as the counterpart of the two well-known triplets of zero modes that arise in the 
electronic spectra of the fullerenes in the continuum limit. Beyond that similarity, 
it is clear that the results from our effective Dirac theory are a particular 
realization of the correspondence between the number of zero modes of the Dirac 
equation and the value of the gauge flux traversing the two-dimensional space.

With our numerical investigation of the electronic spectra of graphene wormholes, 
we have seen that these may fall into two different classes, depending on whether 
the $(6n,0)$ nanotube bridge corresponds to $n$ being a multiple of 3 or not. The 
numerical computations performed in the tight-binding approach have shown that
two triplets of localized states appear in the wormholes where $n$ is equal to a 
multiple of 3, while in the rest of the cases there are two localized states 
at very low energies. This turns out to be in agreement with the analysis 
establishing that the effective gauge flux from a combination of topological 
defects depends in general on their relative location. Thus, only when the 
separation between two heptagonal carbon rings is given by a vector $(N,M)$ 
(in units of the translation vectors of the hexagonal lattice) such that $N-M$ 
is a multiple of 3, the effective flux from the pair of heptagons corresponds 
to the sum of the individual fluxes\cite{lc}. 
This count can be used to show that the 
predictions of the number of localized modes from the effective Dirac theory 
are consistent with the two different classes of electronic spectra displayed 
in the numerical diagonalization of the tight-binding hamiltonians. 

In this paper we have focused on the analysis of graphene wormholes where the
heptagonal carbon rings are regularly distributed around each end of the 
nanotube bridge. Such a constraint leads to wormhole geometries  
in which the graphene branches are connected by zig-zag $(6n,0)$ or
armchair $(6n,6n)$ nanotubes. This can be already realized from the study of
the carbon nanotube-graphene junctions carried out in Ref. \cite{ngj}.
We have checked that the graphene wormholes with armchair bridge not 
considered here fall actually in the same class mentioned above with the two 
triplets of localized states at very low energy. This is again consistent
with the rule giving the effective gauge flux of a combination 
of topological defects, which leads to the maximal value for a wormhole with
$(6n,6n)$ bridge as the separation between heptagonal rings at each 
end is given then by a vector $(n,n)$. It is also possible to think of
more general graphene wormholes in which the topological defects do not 
have a regular distribution. In these cases, we expect anyhow that the 
number of localized states will be dictated again by the effective gauge
flux going through the two-dimensional space. 

For practical purposes, the most significant achievement of our investigation 
has been to show that the low-energy electronic properties of the graphene 
wormholes are consistently described by an effective theory of two Dirac fermion 
fields in the curved geometry of the wormhole. This may open the way to use 
real samples of the curved carbon material as a playground to experiment with 
the interaction between the background curvature and the Dirac fields. In this 
regard, we recall that carbon nanotube-graphene junctions have been already 
produced in the laboratory\cite{fu},
and it is also conceivable that graphene wormholes may be fabricated starting
from the easily available graphene bilayers. The investigation of the 
condensed matter system can make then possible to study relevant gravitational 
effects related to the Dirac character of the electron quasiparticles, which
otherwise would be only accessible at the much higher energies typical of 
the astrophysical phenomena.

\section{Acknowledgements}
We thank F. Barbero, F. Guinea and E. J. S. Villase\~{n}or for very fruitful 
discussions. The financial support from MICINN (Spain) through grant
FIS2008-00124/FIS is gratefully acknowledged.



\end{document}